\documentclass[particles,article,accept,moreauthors,pdflatex,10pt,a4paper]{mdpi}

\usepackage{booktabs}
\usepackage{multirow}
\usepackage{soul}
\usepackage{microtype}

\def\dac{\displaystyle\frac}
\def\[{\left[}
\def\]{\right]}
\def\({\left(}
\def\){\right)}
\usepackage{amsmath}
\usepackage{amssymb}
\usepackage{graphicx}
\firstpage{1}
\articlenumber{x}
\doinum{10.3390/------}
\pubvolume{2}
\pubyear{2018}
\copyrightyear{2018}
\history{Received: 27 January 2018; Accepted: 22 February 2018; Published: date}

\Title{Limits on the Reconstruction of a Single Dark Energy Scalar Field Potential from SNe Ia Data}

\Author{{Arpine Piloyan} $^{1,2}$*{$^{,\dagger}$}, Sergey Pavluchenko $^{2,\dagger}$\orcidA{} and Luca Amendola $^{3,\dagger}$}

\AuthorNames{Arpine Piloyan, Sergey Pavluchenko and Luca Amendola}

\address{%
$^{1}$ \quad A. I. Alikhanian National Science Laboratory, Yerevan Physics Institute, Alikhanyan broth. 2, {0036 Yerevan},~Armenia\\
$^{2}$ \quad Programa de P\'os-Gradua\c{c}\~ao em F\'isica, Universidade Federal do Maranh\~ao (UFMA), \mbox{S\~ao Lu\'is {65085-580}, Maranh\~ao, Brazil}; sergey.pavluchenko@gmail.com\\
$^{3}$ \quad Institut F\"ur Theoretische Physik, Ruprecht-Karls-Universit\"at Heidelberg, Philosophenweg 16, 69120~Heidelberg, Germany; l.amendola@thphys.uni-heidelberg.de
}

\corres{Correspondence: {piloyanarpine@yahoo.com; Tel.: +374-10-352-041}}

\firstnote{These authors contributed equally to this work.}
\abstract{In this paper we perform a reconstruction of the scalar field potential
responsible for cosmic acceleration using SNe Ia data. After describing
the method, we test it with real SNe Ia data---Union2.1
and JLA SNe datasets. We demonstrate that with the current data precision
level, the~full reconstruction is not possible. We discuss the problems
which arise during the reconstruction process and the ways to overcome
them.}

\keyword{cosmology; Dark Energy; scalar field; SNe Ia data}

\begin{document}
%%%%%%%%%%%%%%%%%%%%%%%%%%%%%%%%%%%%%%%%%%
%% Only for the journal Gels: Please place the Experimental Section after the Conclusions

%%%%%%%%%%%%%%%%%%%%%%%%%%%%%%%%%%%%%%%%%%

\section{Introduction}

Since the discovery of the accelerated expansion of the universe in
1998 \cite{riess,perl}, scientists face new
problems in fundamental physics, since general relativity (GR) cannot accommodate
accelerated expansion without a cosmological constant, whose theoretical
difficulties are well-known~\cite{carrol, pad, ss00}.
Many theories and methods have been proposed to solve this puzzle.
Among them, we mention
phantom cosmology~\cite{rev_add1}, tachyonic matter~\cite{rev_add2}, braneworld scenarios~\cite{rev_add3},
scalar-tensor theories~\cite{rev_add4}, f(R)-gravity~\cite{rev_add5, rev_add5.1}, holographic gravity~\cite{rev_add6, rev_add7}, Chaplygin gas~\cite{rev_add8, rev_add9, rev_add10, rev_add11},
models with extra (i.e., more than four) dimensions~\cite{rev_add12}, neutrinos of varying mass~\cite{rev_add13, rev_add14} and many others. In addition, there are also models with fluid-based approaches (e.g.,
~\cite{rev_add14.5}) or
Cardassian cosmology~\cite{rev_add15, rev_add16, rev_add17}.
The~next step, naturally, is to find
the true theory which will best explain the experimental data. To
do this, we must start to look for it in the most general way. One
of the analytical general approaches is Horndeski's theory ~\cite{horndeski},
which builds the most general ghost free Lagrangian scalar field.
This class of models is however too large to be reconstructed without
{a priori} assumptions. This kind of work has been done before;
however, in most cases, the previous attempts have employed a parameterization
with a small number of free parameters (see, e.g.,~\cite{starobinsky, ellis, turner, saini}).
There are also many model independent attempts for the Dark Energy reconstruction; some of them can be found the the following publications~\cite{add1, add2, add3, add4, add5, add6, add7, add8, add9, add10}.
In this paper,
we employ instead a fully non-parametric reconstruction of the scalar
field potential. Unfortunately, our conclusion is that the present
accuracy of data is insufficient to allow for such a reconstruction. It is worth mentioning another non-parametric reconstruction method, which is, again, not fully general but depends on the choice of covariance function, the so-called Gaussian process (see~\cite{saikel}).

The structure of the manuscript is as follows: in Section  2, we present the principle and formulae for the Dark Energy potential reconstruction from the supernovae type Ia (SNe Ia) data. The main data analysis and reconstruction as well as the
  errors analysis is performed in Section 3. Discussion and Conclusions, where we discuss the reasons and a way to improve the method, are provided correspondingly in Sections 4 and 5. We show that the lack of statistical data and accuracy does not allow us to apply the method effectively. However, with some improvement and better data, this~analysis may be irreplaceable because of its generality.
  %NOTE: Sections not in \ref{} format.
  We used in this work only Ia type SNe data as a first step, however, baryonic acoustic oscillations (BAO) as well as perturbations  using the cosmic microwave background (CMB) data  can also be added. These will be taken into account in our next~papers.

\section{Potential Reconstruction Technique}

As we mentioned in the Introduction, we restore the scalar
field potential in the most general form, thus no assumptions and priors
are made about its functional form. Then, the procedure of the recovery
is as follows: as input data, we have distance modulus $\mu_{i}$
with its error $\delta\mu_{i}$ for each $i$th supernova at redshift
$z_{i}$ (possibly with error $\delta z_{i}$, depending on the data).
To smooth the individual values of the distance modulus, we bin the
supernovae. We use two main binning methods: binning
with equal $\Delta z$ and with equal $\Delta N$ (number of supernovae),
as well as their combination, if necessary.
It is worth mentioning that we have used the maximum number in bin such as more number of SNe in each bin does not improve the result and keeps the statistic acceptable.
Both these methods have
their advantages and disadvantages, and in due time we shall
discuss them.

Assuming Gaussian nature of the SNe errors, we determine the mean
value for the distance modulus in each bin as an average over the
individual values $\overline{\mu}_{j}=(\sum_{i}\mu_{i})/N_{j}$ and
the error as $\sigma_{\overline{\mu}_{j}}=\sqrt{(\sum_{i}\sigma_{i}^{2})/N_{j}}$,
where $N_{j}$ is the number of supernovae in $j$th bin. We also
define $\delta z$---the error in $z$---as the half-width
of the bin (in the case we have $\delta z_{i}$ for each individual supernova,
the definition for the error in binned $z$ becomes more complicated).
For equal $z$ binning, it gives the same value for $\delta z$ but,
for alternative binning, it will be different, so we keep this notation
for the general case.

The process of binned data is as follows: first we transform the distance
modulus and its error into the comoving distance $D_{M}$:
\begin{equation}
\begin{array}{l}
D_{M}=\dac{10^{{\mu/5 + 1}}}{1+z},~~\delta D_{M}=\delta\(\dac{10^{\mu/5 + 1}}{1+z}\)=\cdots=D_{M}\dac{\ln10\delta\mu}{5}+\dac{\delta z}{(1+z)}.
\end{array}\label{dL}
\end{equation}

We proceed further to $H(z)$, which is defined as $H(z)={(dD_{M}/dz)}^{-1}$.
In practice, for simplicity and to avoid losing too much data, we use
one-step differentiation scheme. Then, the value and the error take
the form
\begin{equation}
\begin{array}{l}
H(z)=\dac{1}{D_{M}'},~~\delta H(z)=\delta\({\dac{1}{D_{M}'}}\)=\dac{\delta D_{M}'}{(D_{M}')^{2}},~~D_{M}'=\dac{D_{M,2}-D_{M,1}}{z_{2}-z_{1}},\\
\\
\delta(D_{M}')=\delta\({\dac{D_{M,2}-D_{M,1}}{z_{2}-z_{1}}}\)=\cdots=\dac{\delta D_{M,2}+\delta D_{M,1}}{z_{2}-z_{1}}+|D_{M}'|\dac{\delta z_{2}+\delta z_{1}}{z_{2}-z_{1}}.
\end{array}\label{Hz}
\end{equation}

For equal $z$ binning, all $\delta z$ are equal to each
other, but, in the general case, this is not true, so we keep the general
form of the errors expression. Similar to the procedure described
above, we calculate the derivative of $H(z)$ with respect to $z$:
\begin{equation}
\begin{array}{l}
H'=\dac{H_{2}-H_{1}}{z_{2}-z_{1}},~~\delta H'=\cdots=\dac{\delta H_{2}+\delta H_{1}}{z_{2}-z_{1}}+|H'|\dac{\delta z_{2}+\delta z_{1}}{z_{2}-z_{1}}.
\end{array}\label{dH}
\end{equation}

Now, with both $H(z)$ (see Equation~(\ref{Hz})) and $H'$ (see Equation~(\ref{dH}))
calculated, we can recover the potential $V(z)$ and the kinetic part
$(d\phi/dz)^{2}$. The equations for them and their errors are the
following:

\begin{equation}
\begin{array}{l}
\tilde{V}\equiv\dac{8\pi G}{3H_{0}^{2}}V(z)=\dac{H(z)^{2}}{H_{0}^{2}}-\dac{H(z)H'(z)(1+z)}{3H_{0}^{2}}-\dac{\Omega_{m}^{0}(1+z)^{3}}{2},\\
\\
\delta\tilde{V}=\dac{2H\delta H}{H_{0}^{2}}+\dac{(1+z)H'\delta H+H(1+z)\delta H'+HH'\delta z}{3H_{0}^{2}}+\dac{3(1+z)^{2}\Omega_{m}^{0}\delta z}{2};
\end{array}\label{Vz}
\end{equation}

\begin{equation}
\begin{array}{l}
\tilde{\({\dac{d\phi}{dz}}\)^{2}}\equiv\dac{8\pi G}{3H_{0}^{2}}{\dac{d\phi}{dz}}^{2}=\dac{2H'(z)}{3H(z)H_{0}^{2}(1+z)}-\dac{\Omega_{m}^{0}(1+z)}{H^{2}},
\\ \\
\delta{\tilde{\({\dac{d\phi}{dz}}\)^{2}}}=\dac{2}{3H_{0}^{2}}
\dac{\delta H'}{H(1+z)}+\dac{H'\delta H}{H^{2}(1+z)}+\dac{H'\delta z}{H(1+z)^{2}}+\Omega_{m}^{0}{\dac{\delta z}{H^{2}}+\dac{2(1+z)\delta H}{H^{3}}}.
\end{array} \label{dfdz2}
\end{equation}

Equations (\ref{Vz}) and (\ref{dfdz2}) give us the dimensionless
potential and the kinetic term for the scalar field, $\tilde{V}(z)$ and $(d\tilde{\phi}/dz)^{2}$, respectively;
the latter (if positive) could be integrated to get $\phi(z)$ (with
an additive constant). Finally, eliminating $z$ from $\tilde{V}(z)$
and $\tilde{\phi}(z)$ in Equations (\ref{Vz}) and (\ref{dfdz2}),
we reconstruct $\tilde{V}(\tilde{\phi})$ (again, up to additive constant).
For simplicity, we use the rectangle method for integration,
so~the error propagation for the remaining steps is as follows:
\begin{equation}
\begin{array}{l}
\dac{d\phi}{dz}=\sqrt{(\dac{d\phi}{dz})^{2}}\Rightarrow\delta{\dac{d\phi}{dz}}=1/2 \times \delta{\({\dac{d\phi}{dz}}\)^{2}}/{\dac{d\phi}{dz}};\\
\\
\phi=\int{\dac{d\phi}{dz}}dz\cong{\dac{d\phi}{dz}}|_{z_{central}}\Delta z\Rightarrow\delta\phi=\Delta z\times\delta{\dac{d\phi}{dz}}.
\end{array}\label{rest}
\end{equation}

It is worth mentioning that the described procedure is quite similar to the approach employed by the cosmography (see~\cite{cosmo1} for review;~\cite{cosmo2, cosmo3, cosmo4, cosmo5} for the effects on GR level;~\cite{cosmo6} for extended gravity; \cite{cosmo7} for $f(R)$ gravity; and \cite{cosmo8, cosmo9} for $f(T)$ gravity).
%NOTE: Please confirm
In turn, this is quite close to $\omega$CDM approach and we discuss the similarities of these methods in the
 Discussion Section.

\section{Reconstruction Procedure by Using Real SNe Ia Data}

\begin{figure}[H]
\centering
\includegraphics[width=1\textwidth]{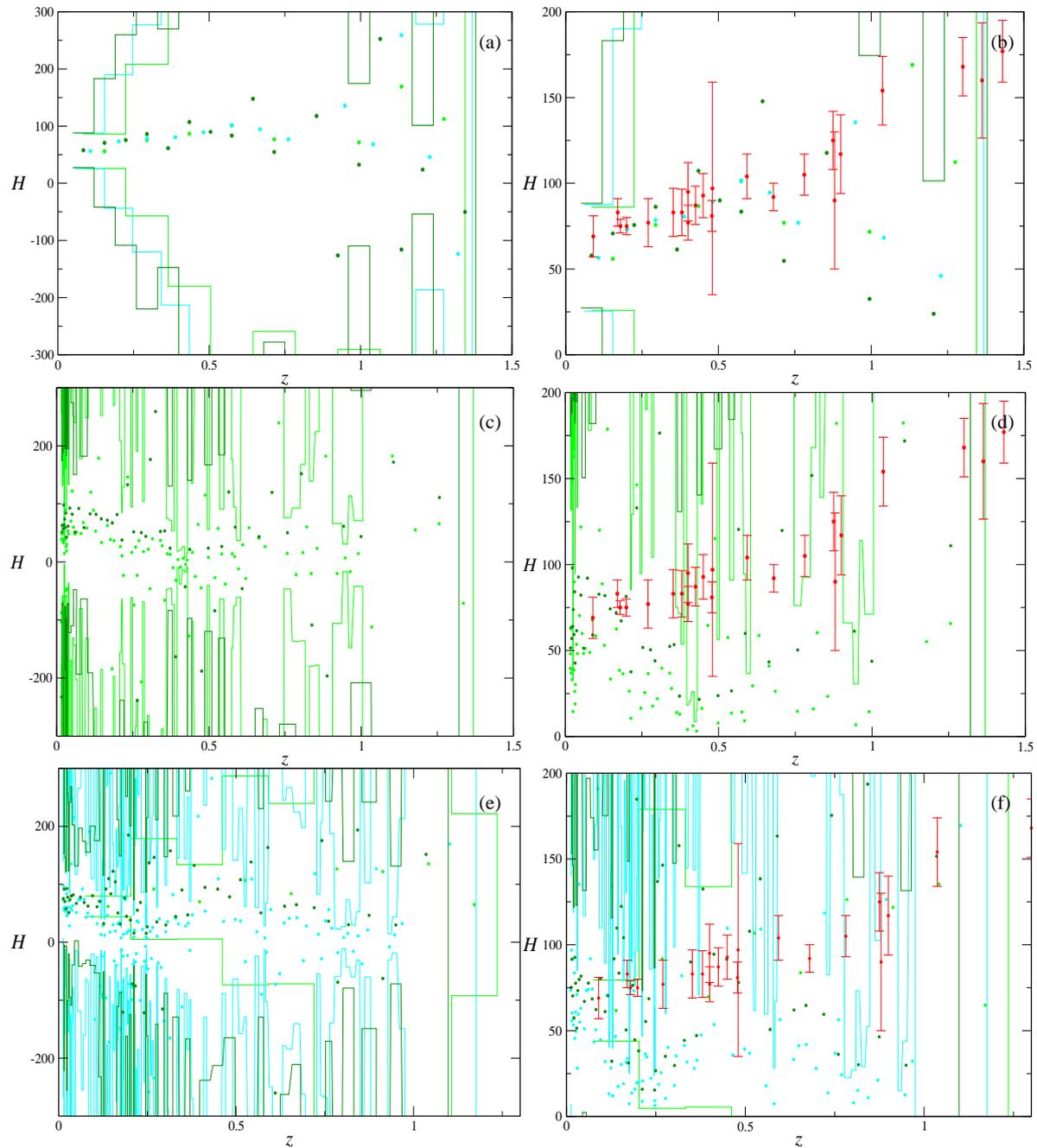}
\caption{Results for $H(z)$: areas of $\pm1\sigma$ for each bin and the central
values for the following cases and datasets: ({\bf a},{\bf b}) Equal-$z$ binning of the Union2.1 data---10 bins in green,
15 bins in cyan and 20 bins in dark green---for ``large-scale'' data
({\bf a}); and physically-significant range ({\bf b}). ({\bf c},{\bf d}) Equal-$N$ binning of the Union2.1 data---$N=5$ in green and $N=10$ in dark green---for ``large-scale'' data ({\bf c}); and physically-significant range ({\bf d}).
 ({\bf e},{\bf f})
 Equal-$N$ with 10 bins in green and equal-$N$
binning of the JLA data---$N=5$ in cyan and $N=10$ in
dark green---for ``large-scale'' data ({\bf e}); and physically-significant
range ({\bf f}). Additionally, in ({\bf b},{\bf d},{\bf f}), we added
non-SNe $H(z)$ data points with errors in red (see text for more
details).}
\label{fig2}
\end{figure}

We use two existing SNe Ia sets: the 580 SNe from Union2.1
and the 740 JLA SNe. In addition, we use two different binning
techniques, with equal $\Delta z$ and equal $\Delta N$
(number of data points per bin), and compare the results. First, we
examine Union2.1 data. We bin data over 10, 15 and 20 bins with equal
$z$. With 10 bins, the minimal number of SNe per bin is 9; for 15 bins,
it is 5; and, for 20 bins, it is 4. If we consider binning into 25 bins,
the minimal number of SNe per bin would be~2, which is too low from
the statistical point of view, so we do not consider binning into
25 or more bins. Using the above described method, we recover everything
up to $V(z)$ and $(d\phi/dz)^{2}$; as $H(z)$ is the quantity that is
directly probed by observations, we present it as an intermediate
step in Figure~\ref{fig2}a--f. There (and in the plots
to follow) we put $\pm\sigma$ area contours and the central values
with the same color. Different panels and different colors correspond
to different cases: in Figure~\ref{fig2}a,b, we presented $H(z)$ for
equal-$z$ binning of Union2.1 data---10 bins in green,
15 bins in cyan and 20~bins in dark green. Figure~\ref{fig2}a, we present
the ``large-scale'' structure, and, in Figure~\ref{fig2}b, a more realistic range.
We~also added the $H(z)$ values with errors from SNe-independent
data (from chronometers, etc.---see~\cite{moresco}) in
red. One~can see that, up to $z\sim0.5$, they are more or less consistent
but at larger $z$, the consistency is lost and we even have negative
values for $H(z)$ (in the cyan and dark green lines).

Figure~\ref{fig2}c,d
 correspond to equal $N$ binning of the Union2.1 data.
The~green points correspond to $N=5$ binning while dark green
to $N=10$. We plot ``large-scale'' data in Figure~\ref{fig2}c and more realistic
range in Figure~\ref{fig2}d. One can immediately see that the final result is very
much scattered and contain very little information. This is because
 we have a lot of data points at low $z$ and they create
many bins there. With binning to just 5 and 10 SNe, the effect of
averaging is wearing off and so the individual uncounted systematics
for each SNe start to manifest themselves. On the contrary, if we
choose large $N$, we will have fewer bins in high $z$ range, which
is the range of our interest, thus we will lose important
data. Overall, the case with $N=5$ and $N=10$ do not allow any physical
conclusion.

Finally, in Figure~\ref{fig2}e,f, we present the data
for JLA SNe---equal-$z$ with 10 bins in green, equal-$N$
with $N=5$ in cyan, and $N=10$ in dark green. For equal-$z$, we have
only one sample due to the fact that even for just 10 bins the minimal
number of SNe per bin is 2, and with 15 bins there are bins with no
SNe at all, so we decided to consider only this case. In Figure~\ref{fig2}e,f,
one can see that equal-$N$ sets are as messy as in Union2.1 case,
making equal-$N$ binning less favorable than equal-$z$~one.

The result for $H(z)$ clearly demonstrate that the reconstruction
of $V(\phi)$ is hopeless. Notice that, for restoration, we require
both $H(z)$ and its derivative $H'$, which is expected to be even
more noisy.

\begin{figure}
\centering
\includegraphics[width=0.98\textwidth]{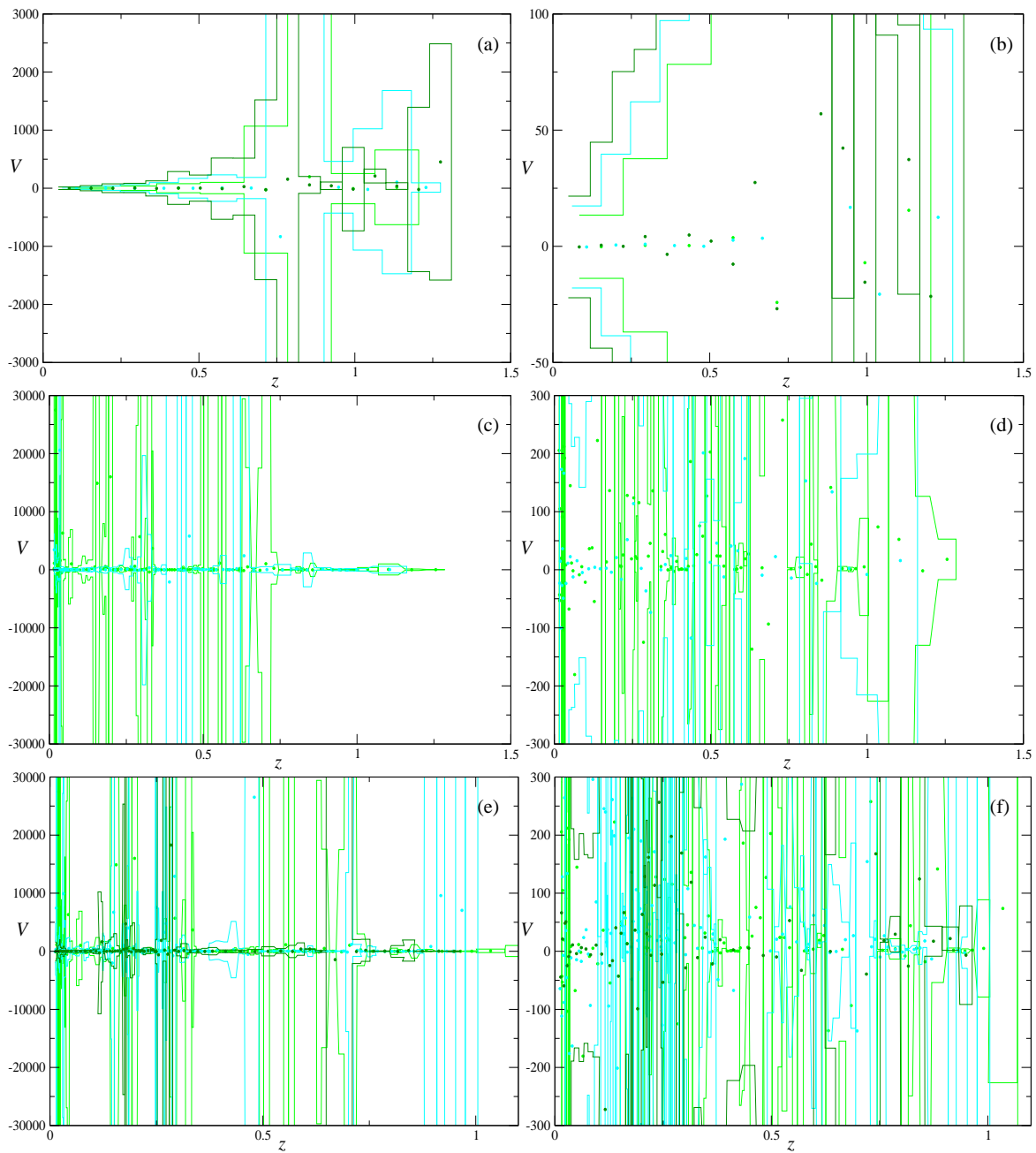}\vskip-1.5mm
\caption{{Results}
for $V(z)$ recovering:
areas of $\pm1\sigma$ for each bin and the central
values for the following cases and datasets: ({\bf a},{\bf b}) Equal-$z$ binning of the Union2.1 data---10 bins in green,
15 bins in cyan and 20 bins in dark green---for ``large-scale'' data
({\bf a}); and physically-significant range ({\bf b}). ({\bf c},{\bf d}) Equal-$N$ binning of the Union2.1 data---$N=5$ in green and $N=10$ in dark green---for ``large-scale'' data ({\bf c}); and physically-significant range ({\bf d}).
 ({\bf e},{\bf f})
 Equal-$N$ with 10 bins in green and equal-$N$
binning of the JLA data---$N=5$ in cyan and $N=10$ in
dark green---for ``large-scale'' data ({\bf e}); and physically-significant
range ({\bf f}).
 (see the text for more details).}
\label{fig3}
\end{figure}

\begin{figure}
\centering
\includegraphics[width=1\textwidth]{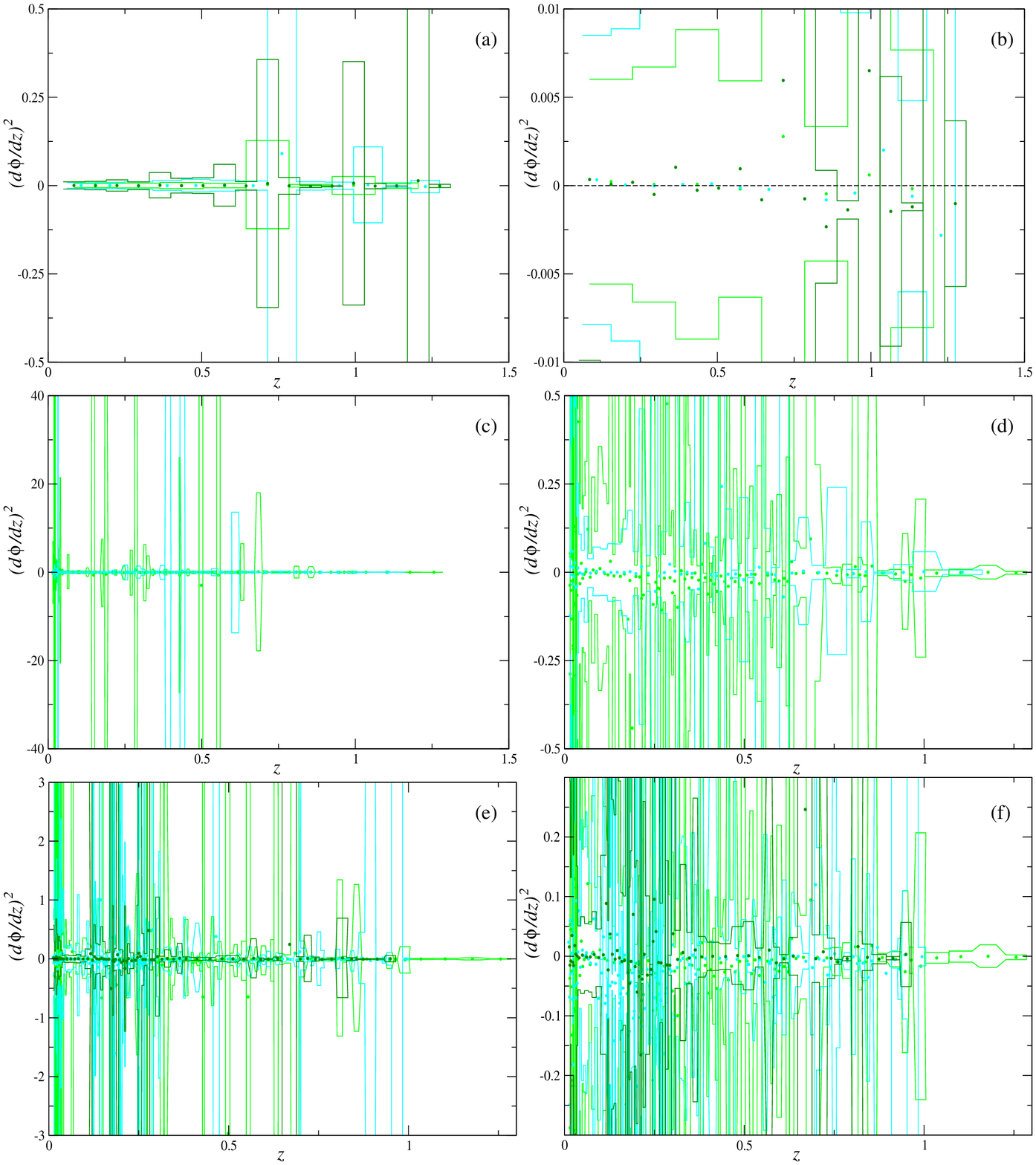}
\caption{{Results} for $\({\dac{d\phi}{dz}}\)^{2}$ recovering:
areas of $\pm1\sigma$ for each bin and the central
values for the following cases and datasets: ({\bf a},{\bf b}) Equal-$z$ binning of the Union2.1 data---10 bins in green,
15 bins in cyan and 20 bins in dark green---for ``large-scale'' data
({\bf a}); and physically-significant range ({\bf b}). ({\bf c},{\bf d}) Equal-$N$ binning of the Union2.1 data---$N=5$ in green and $N=10$ in dark green---for ``large-scale'' data ({\bf c}); and physically-significant range ({\bf d}).
 ({\bf e},{\bf f})
 Equal-$N$ with 10 bins in green and equal-$N$
binning of the JLA data---$N=5$ in cyan and $N=10$ in
dark green---for ``large-scale'' data ({\bf e}); and physically-significant
range ({\bf f}).
(see the text for more details).}
\label{fig4}
\end{figure}

The results for $V(z)$ for the same cases are presented in Figures~\ref{fig3}
and \ref{fig4}. Both figures keep the same designations as in Figure~\ref{fig2}: Figures~\ref{fig3}a,b
and \ref{fig4}a,b correspond to Union2.1 SNe
with equal-$z$ binning; Figures~\ref{fig3}c,d
and \ref{fig4}c,d to Union2.1
with equal-$N$ binning; and Figures~\ref{fig3}e,f
and \ref{fig4}e,f to JLA SNe. The
color distribution also follow that in Figure~\ref{fig2}.

\unskip

\begin{figure}
\centering
\includegraphics[width=1\textwidth]{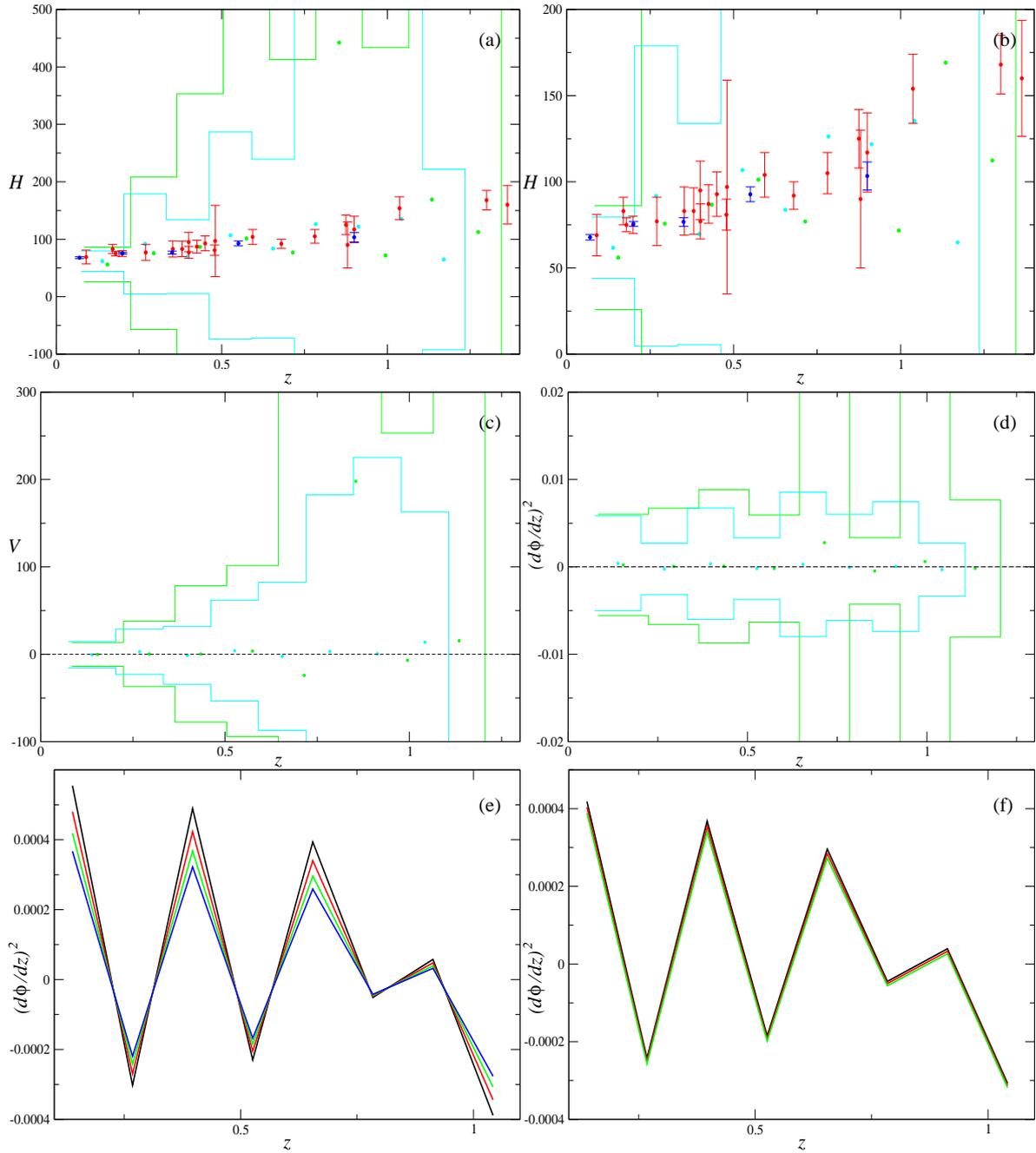}
\caption{Combined
results for equal-$z$ binning with 10 bins (green points
correspond to Union2.1 and cyan to JLA data):  ({\bf a},{\bf b}) $H(z)$
with non-SNe data from~\cite{moresco} as red points and data
from~\cite{riess17} as blue points;  ({\bf c}) recovered $V(z)$;
 ({\bf d}) recovered $\({\dac{d\phi}{dz}}\)^{2}$ ; ({\bf e}) the
effect of the variation of $H_{0}$;  and ({\bf f}) the effect of the variation $\Omega_{m}$
 (see the text for more details).}
\label{fig5}
\end{figure}

In Figures~\ref{fig3} and \ref{fig4}, we can clearly see that equal-$N$
binning is extremely noisy, as we predicted above. The only results
which looks more-or-less physical (or at least which does not look
totally unphysical) are the results for equal-$z$ binning. We chose the
two most ``physical'' ones---10 bins from Union2.1 and
JLA datasets and put them together in Figure~\ref{fig5}. In Figure~\ref{fig5}a,b,
 we present $H(z)$ data---green for Union2.1
and cyan for JLA datasets. We also added non-SNe $H(z)$ data points
from~\cite{moresco} (red circles with error bars) and $H(z)$ from
SNe from~\cite{riess17} (blue circles with error bars) to compare
our results with them. One can see kind of an agreement between the
data for $z\lesssim0.5$ but for $z>0.5$ $H(z)$ start to become
noisy again. In Figure~\ref{fig5}c, we present the results for $V(z)$ recovering
and, in Figure~\ref{fig5}d, for $\({\dac{d\phi}{dz}}\)^{2}$.
One can see that, even for the chosen two best binning options, we
still have negative central values for $\({\dac{d\phi}{dz}}\)^{2}$,
which makes it impossible to recover $\phi(z)$.

The recovering of $V(z)$ and ${\dac{d\phi}{dz}}^{2}$
was performed for $H_{0}=68$ km/s/Mpc and $\Omega_{m}=0.25$. As one
can see from {Equations}
(\ref{Vz}) and (\ref{dfdz2}), the recovering procedure
use $H_{0}$ and $\Omega_{m}$ as a parameters, so we decided to vary
them to see the influence. It is presented in
Figure~\ref{fig5}e,f. There, we present $\({\dac{d\phi}{dz}}\)^{2}$
recovered for different values of $H_{0}$ and $\Omega_{m}$. The
potential $V(z)$ could be recovered for any values of $H_{0}$ and
$\Omega_{m}$. The expression $\({\dac{d\phi}{dz}}\)^{2}$
, instead, must of course be positive so we focus on it. In Figure~\ref{fig5}e,
 we fixed $\Omega_{m}=0.2$ and varied $H_{0}=60$ km/s/Mpc (black line),
64 km/s/Mpc (red line), 68 km/s/Mpc (green line), and 72 km/s/Mpc (blue line). One can see that
the effect is not big but visible and the values grow with increased
$H_{0}$. In Figure~\ref{fig5}f, we fixed $H_{0}=60$ km/s/Mpc and varied $\Omega_{m}=0.2$
(black line), 0.25 (red line) and 0.3 (green line). One can see that
$\Omega_{m}$ does not practically change the value for $\({\dac{d\phi}{dz}}\)^{2}$.

\begin{figure}
\includegraphics[width=1\textwidth]{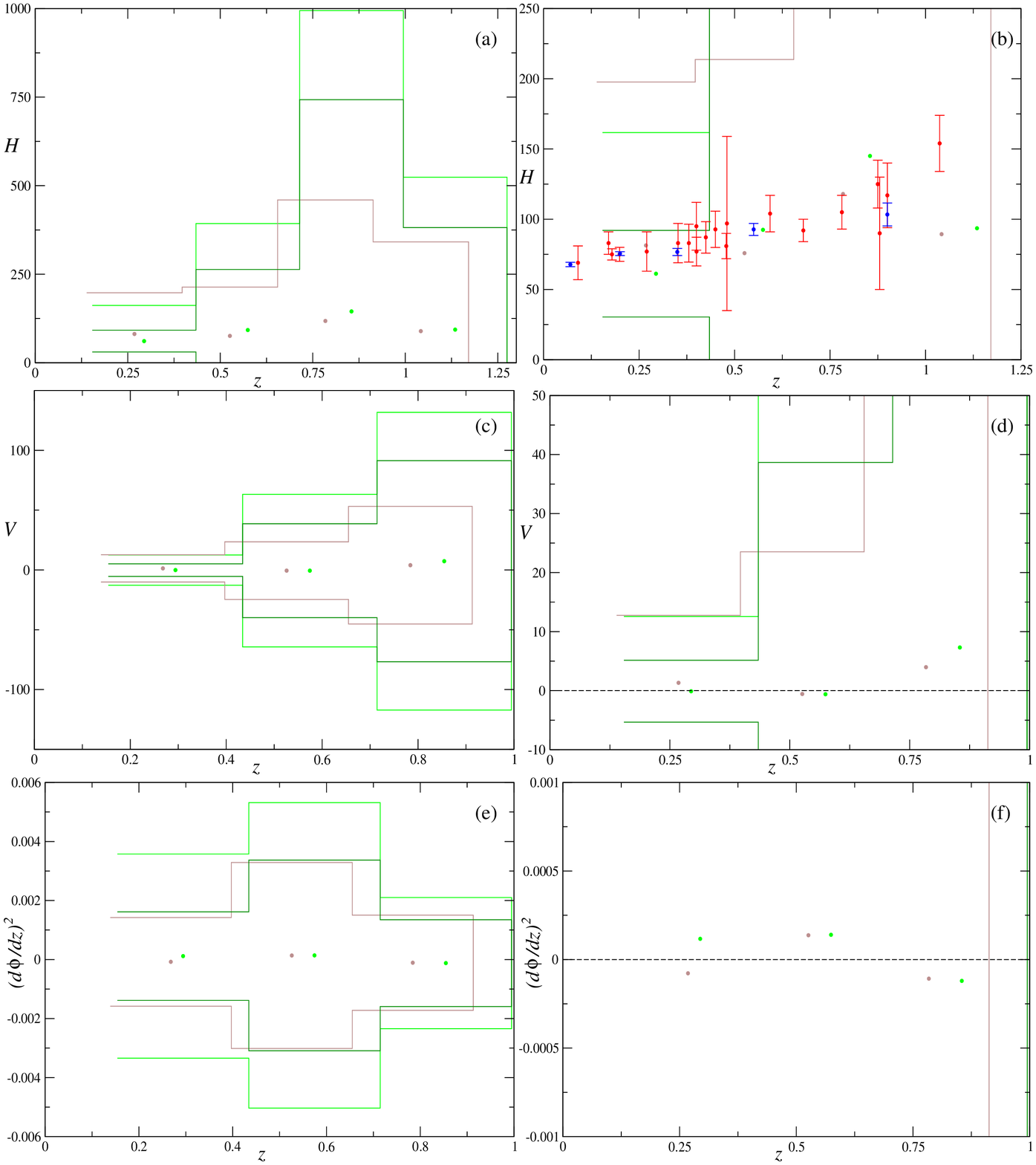}
\caption{Combined results for equal-$z$ binning with five bins (green points correspond
to Union2.1 and brown to JLA data); central values designed as filled
circles while $1\sigma$ error areas are bounded by the line of the
same color:  ({\bf a},{\bf b}) $H(z)$ with non-SNe data  from~\cite{moresco} as red
points and data from~\cite{riess17} as blue points (``large-scale''
({\bf a}); and ``fine'' ({\bf b}));  ({\bf c},{\bf d}) recovered $V(z)$
(``large-scale'' ({\bf c}); and ``fine'' ({\bf d}));  and ({\bf e},{\bf f}) recovered $\({\dac{d\phi}{dz}}\)^{2}$ (``large-scale'' ({\bf e}); and ``fine''
({\bf f})). For Union2.1 data, light green corresponds to $\delta z\protect\ne0$
while dark green to $\delta z=0$ (see the text for
more details).}
\label{fig5bins}
\end{figure}

Finally, we decided to try five bins, as even ten have not given us a
realistic reconstruction. Thus, we split both Union2.1 and JLA datasets
into five equal-z bins and perform exactly the same analysis. This time,
we also check the effect of $\delta z=0$ on the error propagation
in real data. The results are presented in Figure \ref{fig5bins}. There,
in Figure \ref{fig5bins}a--f, we put the central values as filled circles, while $1\sigma$
error areas are bounded by the line of the same color: green for Union2.1
and brown to JLA data. As for the former, the light green corresponds
to $\delta z\ne0$, while dark green to $\delta z=0$. The $H(z)$
data are presented in Figure \ref{fig5bins}a,b with non-SNe data from~\cite{moresco}
as red points and data from~\cite{riess17} as blue points (``large-scale''
in Figure \ref{fig5bins}a and ``fine'' in Figure \ref{fig5bins}b); the recovered $V(z)$ is presented
in Figure \ref{fig5bins}c,d (``large-scale'' in Figure \ref{fig5bins}c and ``fine''
in Figure \ref{fig5bins}d); and the recovered $\({\dac{d\phi}{dz}}\)^{2}$ in Figure \ref{fig5bins}e,f (``large-scale'' in Figure \ref{fig5bins}e and ``fine'' in Figure \ref{fig5bins}f).

One can see that the recovered $H(z)$ data points are mostly in agreement
with non-SNe data---as in the 10-bins case as well. Obviously,
with $\delta z=0$, the errors are much smaller than in the $\delta z\ne0$
case. Unfortunately, the potential and the kinetic term reconstructions
do not improve---we also have some points of the kinetic
term in the negative domain, which prevents the complete reconstruction.

In Figure \ref{fig5bins}b, we can see that even our best $H(z)$
still have an error budget which is several times bigger then for
the existing $H(z)$ data---non-SNe data from~\cite{moresco}
as well as recent SNe-based $H(z)$ data from~\cite{riess17}. We
are using 26 $H(z)$ non-SNe data points ({{we were using 26 data points, but~at the time there could be found
maximum of 36 non-SNe data points (chronometers, BAO etc.), see~e.g.,}~\cite{d.wang},
{but~one need to carefully investigate the nature of these data, their
quality and errors before usage}})
---out of 30 from~\cite{moresco} we dropped four from~\cite{zhang},
since their precision is much less then the rest of the data from~\cite{moresco}.

\section{Discussion}

In this paper, we attempt to reconstruct the scalar field
potential, which could be responsible for Dark Energy. In fact, there
are a lot of such attempts (see, e.g.,~\cite{saini, add7, add8, add9, add10}), but with one
crucial difference---all of them using parameterization
for either $H(z)$ or the effective equation of state, or some other
physical quantity.
One of such parameterizations is $\omega$CDM, where one assumes parameterization of the equation of state of the Dark Energy in the form \mbox{$\omega=\omega_{0}+\omega_{1}(1+z)+\ldots$}.
In the simplest case, $\omega = \omega_0 \ne -1$, the Dark Energy could be described as scaling~\cite{sc1, sc2} or tracker~\cite{tr} solutions---exponential, inverse power-law and several others~\cite{tr}.
With more terms in the $\omega$ decomposition, the solutions for the scalar field becomes less likely to be found, especially in the closed form, which makes the connection of $\omega$CDM with scalar fields less obvious.
Clearly, the potential resulting from $\omega$CDM is a simple exponential or power-law potential that depends only on one or two parameters. In this case, the reconstruction will be much less noisy than in our
parameter-free case.

Another approach is the one used in cosmography (see~\cite{cosmo1} for review and~\cite{cosmo2, cosmo3, cosmo4, cosmo5, cosmo6, cosmo7, cosmo8} for applications)---there, one uses the decomposition on the derivatives of $H(z)$---a similar approach is used in $\omega$CDM but with decomposition over $\omega(z)$. Formally, both cases allow reconstruction of $H(z)$ from the observational (SNe Ia, in particular) data. Despite quite similar procedures
(the reconstruction os the $H(z)$ and then its differentiation to obtain higher derivatives), there are differences---for instance, for our case $H'(z)$ is enough, while, in general, in cosmography and $\omega$CDM higher
derivatives are involved. In this regard, the approaches are also different---we saw that, in our case, without any {\it {Ansatz}} %Is the italics necessary?
for $\mu(z)$ or $H(z)$ or the equation of state or anything else, the reconstructed $H(z)$
is quite noisy, which makes its first derivative $H'$ even more unprecise. If we continue this procedure to higher derivatives, the results will be totally unusable. However, in this procedure, we keep the reconstructed values
totally unbiased. On the contrary, for $\omega$CDM, one introduces the {\it Ansatz} (and, through it, a bias) and reconstructs the $H(z)$ in the reduced (and degenerate) way, as well as its derivatives. In this regard, these approaches are meant for different purposes---one aims at reconstructing the scalar field potential in the most unbiased way, and the other at reconstructing of the $H(z)$ with the least errors. This could be easily seen in Figure~\ref{fig5bins}b: there, one can find our recovered $H(z)$ data points (green and brown) as well as those obtained with non-SNe data:
cosmic chronometers, BAO etc.---\cite{moresco} compilation, red points as well as SNe-based~\cite{riess17} data, blue points. One can see that the central values are in a good agreement up to $z \lesssim 0.75$, for greater
$z$ our data points become messy, but our error bars are severalfold larger. This is the effect of the generality---with no {\it ansatz} for the dynamical variables, we are left with huge errors. On the contrary,
Riess et al. \cite{riess17} use six-parameter parameterization for $H(z)$ resulting in much less errors.

Another problem is that the majority of the data points are located
within $z<0.2$. This is natural---with growth of $z$,
the distance is increasing and so we can observe only most luminous
SNe which affect their number. Here comes the first of the effects
we faced while recovering $H(z)$---the effect of binning.
We used two types of the binning: equal-$N$ binning (so that to have
equal number of SNe per bin) and equal-$z$ binning (equal $z$ width
of each bin). Both of the types have their grounds---with
equal number of supernovae per bin, equal-$N$ binning formally provide
more statistically viable data. However, due to inhomogeneous distribution
of the SNe over $z$, it leads to a lot of bins in low-$z$ region
and only a few in high-$z$. As in reconstruction we are also interested
in high-$z$ region, this scheme is less favorable for us.
Since we are unable to reconstruct the potential, it is hard to judge to effect of binning on the final results, but we can do it for the previous step---namely, reconstruction of the $H(z)$.
The results suggest that for equal-$N$ binning with all considered $N$ the resulting reconstructed $H(z)$ looks quite messy (see Figure~\ref{fig2}c--f). One can see that the reconstructed $V(z)$ (Figure~\ref{fig3}c--f)
and the kinetic term (Figure~\ref{fig4}c--f) are even more messy---the differentiation of the unprecise $H(z)$ only decrease the precision of the result.
The reason behind it lies in huge variability between
the individual bins---the smoothness of the data is not enough to perform such binning. In this regard, if we increase number of SNe per bin, the smoothness increases, but due to highly uneven distribution of the SNe over $z$,
the~bin at highest possible $z$ will be located at $z<1$, which is too low for our initial aim---reconstruction of the Dark Energy scalar field potential.
Thus, we decided that the equal-$N$ binning is not suited for our
cause and more in-depth investigation was performed with equal-$z$ binning.

On the contrary, equal-$z$ binning have different number of SNe per
bin, which makes each bin to have different statistical weight, but this scheme allows us to have sufficient number of data points in the high-$z$ range.
Our analysis demonstrated that with  the increase of bins number the resulting $H(z)$ becomes more messy (see Figure~\ref{fig2}a,b). This is because, at high $z$, there are much less SNe, thus we have
 much less SNe per bin and the impact of each SNe (and so from the individual uncertainties) increase drastically. For the considered SNe datasets, consideration of even 30 bins per range for JLA
 is unreasonable, as in this case there appears empty bins. For 25 bins for JLA the minimal number of SNe per bin is just two, which is too small for statistical significance, so that during primary investigation
we considered 10, 15 and 20 bins per range. The results suggest that
only 10 bins gives more-or-less physical
results (see {Figures~\ref{fig2}a,b, \ref{fig3}a,b, \ref{fig4}a,b and \ref{fig5}a,b}), but still fail to reconstruct the potential, so
that we obviously
need more SNe per bin. Even with as low as five bins (which give
four bins for $H(z)$---one bin less from numeric differentiation,
thus three bins for $H'(z)$ for the same reasons), the reconstruction looks better (see Figure~\ref{fig5bins}) but still not good enough for the successful reconstruction of $V(\phi)$.
Overall, one can see that with decrease of the bins number, the agreement between our reconstructed $H(z)$ and those from other sources (red and blue data points in Figures~\ref{fig2}--\ref{fig5bins} panels (a) and (b))
improves---from around $z \lesssim 0.6$ for 10 bins to $z \lesssim 0.75$ for 5 bins.
 For the JLA data even with 10 bins the minimal
number of SNe per bin is 2, making this set less favorable for our
analysis.

Apart from the binning, it is clear that a major source of the uncertainties---and this time extrinsic uncertainties---is the numerical differentiation. We use it twice---first for $H(z)$ and then for $H'(z)$. With a parametrized $H(z)$
the error propagation would be much simpler and the resulting errors
would be much smaller, but, as we mentioned earlier, this would arbitrarily
restrict {a priori} the family of solutions.

Our results for $H(z)$ reconstruction demonstrate good agreement
with non-SNe-based $H(z)$ from~\cite{moresco} and SN-based $H(z)$
from~\cite{riess17} within $z\lessapprox 0.6$ for both 10 bins case
(see Figure~\ref{fig5}b) and five bins (see Figure~\ref{fig5bins}b),
but with higher $z$ they become very noisy. This leads to more inaccurate
reconstruction of the $H'(z)$ and, as a consequence, quite messy
reconstruction of both $V(z)$ and the kinetic term---both of them enter negative domain, which prevents further reconstruction
of the potential in the $V(\varphi)$ form.

We also investigated the effect of the $H_{0}$ and $\Omega_{m}$
on the reconstructed potential---but even with varying
both $H_{0}$ and $\Omega_{m}$ within more-or-less accepted range,
we failed to find values which lead to the kinetic term reconstruction
with no negative values. This may indicate the problem with the data
(some unaccounted issues with data reduction), with binning, or overestimation
of $H_{0}$ and/or $\Omega_{m}$.

\section{Conclusions}

We have described a scheme for the reconstruction of the potential
of the scalar field, which is supposed to be responsible for the accelerated
expansion. The described scheme is tested with real SNe data available
today---Union2.1 and JLA datasets. We tested two binning
techniques---equal-$N$ and equal-$z$ binning---with different
number of SNe per bin for the former, and different number of bins per  entire $z$ range
for the latter, and find it that only equal-$z$ with number of bins
no more then 10 gives us the reconstruction of $H(z)$ which is in
agreement with independent estimations within $z\lessapprox0.6$.
The reconstructed $V(z)$ and the kinetic term disallow further reconstruction
of the $V(\varphi)$ since the reconstructed kinetic term enters in the negative
domain. Based on the discussion of the previous section, we believe
that the following points, which shall be implemented in the papers
to follow, would improve the results of the reconstruction.

First, it could be useful to find a specific binning---with high enough number of SNe per bin and enough bins per $z$ to
reconstruct all necessary quantities with good precision.
Indeed, as we mentioned, with five bins for $\mu$, we recover only three
bins for the potential, which is quite low number of data
points. This way we could utilize low density of SNe at high $z$
to create as many bins as possible, and wisely split low-$z$ domain
to a number of bins which create as smooth as possible reconstructed
$H(z)$ and its derivative. Of course, this binning will be ``artificial'',
but, as we discussed, it still will be representative and will serve
our cause.

Second, it is possible that the resulting binning will not have bins
equally spaced over $z$, which will increase errors coming from the
numerical differentiation. To counter that, we can forecast the reconstruction
method with future more precise and more extended observations.

Third, we might resort to other methods to deal with reconstruction,
in particular the so-called Gaussian Process (see, e.g.,~\cite{saikel}).
In this method, one assumes that every point in the reconstructed
curve is correlated with every other point with a given correlation
function that depend on a small number of parameters. This correlation
is of course assumed totally by hand but experience with noisy data
show that indeed it leads to a robust reconstruction.

To conclude, usage of more precise data, together with advanced binning
techniques or reconstruction methods, might lead to more realistic Dark
Energy scalar field potential reconstruction.

\vskip6pt
%%%%%%%%%%%%%%%%%%%%%%%%%%%%%%%%%%%%%%%%%%
\acknowledgments{A.P. thanks DAAD and FAPEMA (under project BPV-00040/16) for support. S.P. was supported by FAPEMA under project BPV-00038/16. L.A. acknowledges support from the DFG TR33 “The Dark Universe” project.
}

%%%%%%%%%%%%%%%%%%%%%%%%%%%%%%%%%%%%%%%%%%
\authorcontributions{{The project is set and prior data analysis were made by L.A. and A.P. Whole further improvements of data analysis are made by S.P.. Paper wrote S.P., A.P and L.A.. All the research is done under the guidance of L.A..}}

%%%%%%%%%%%%%%%%%%%%%%%%%%%%%%%%%%%%%%%%%%
\conflictsofinterest{The authors declare no conflict of interest.}

\reftitle{References}

\end{document}